\documentclass{article}
\usepackage{spconf,amsmath,epsfig}
\usepackage{amssymb}
\usepackage{amsmath}
\usepackage{amsfonts}

\title{An ESPRIT-based Approach for Initial Ranging in OFDMA Systems}
\twoauthors
  {Luca Sanguinetti* and Michele Morelli}
{University of Pisa\\
Department of Information Engineering \\Pisa, Italy \\luca.sanguinetti, michele.morelli@iet.unipi.it}
  {H. Vincent Poor}
{Princeton University\\
Department of Electrical Engineering
\\Princeton, NJ USA \\
poor@princeton.edu }
\begin{document}
%
\maketitle\renewcommand\thefootnote{} \footnotetext{*This work was completed
while the author was with Princeton University and it was supported
by the U.S. National Science Foundation under Grants ANI-03-38807 and CNS-06-25637.}
\begin{abstract}
This work presents a novel Initial Ranging scheme for orthogonal
frequency-division multiple-access networks. Users that intend to
establish a communication link with the base station (BS) are normally
misaligned both in time and frequency and the goal is to jointly estimate
their timing errors and carrier frequency offsets with respect to the BS
local references. This is accomplished with affordable complexity by
resorting to the ESPRIT algorithm. Computer simulations are used to assess
the effectiveness of the proposed solution and to make comparisons with
existing alternatives.
\end{abstract}

%
%

\section{Introduction}

\label{sec:intro}

A major impairment in orthogonal frequency-division multiple-access (OFDMA)
networks is the remarkable sensitivity to timing errors and carrier
frequency offsets (CFOs) between the uplink signals and the base station
(BS) local references. For this reason, the IEEE 802.16e-2005 standard for
OFDMA-based wireless metropolitan area networks (WMANs) specifies a
synchronization procedure called Initial Ranging (IR) in which subscriber
stations that intend to establish a link with the BS transmit pilot symbols
on dedicated subcarriers using specific ranging codes. Once the BS has
detected the presence of these pilots, it has to estimate some fundamental
parameters of ranging subscriber stations (RSSs) such as timing errors, CFOs
and power levels.

Initial synchronization and power control in OFDMA was originally discussed
in \cite{Krinock2001} and \cite{Minn2004} while similar solutions can be
found in \cite{Mahmoud2006}-\cite{Zhou2006}. A different IR approach has
recently been proposed in \cite{Minn07}. Here, each RSS transmits pilot
streams over adjacent OFDMA blocks using orthogonal spreading codes. As long
as channel variations are negligible over the ranging period, signals of
different RSSs can be easily separated at the BS as they remain orthogonal
after propagating through the channel. Timing information is eventually
acquired in an iterative fashion by exploiting the autocorrelation
properties of the received samples induced by the use of the cyclic prefix
(CP).

All the aforementioned schemes are derived under the assumption of perfect
frequency alignment between the received signals and the BS local reference.
However, the occurrence of residual CFOs results into a loss of
orthogonality among ranging codes and may compromise the estimation accuracy
and detection capability of the IR process. Motivated by the above problem,
in the present work we propose a novel ranging scheme for OFDMA networks
with increased robustness against frequency errors and lower computational
complexity than the method in \cite{Minn07}. To cope with the large number
of parameters to be recovered, we adopt a three-step procedure. In the first
step the number of active codes is estimated by resorting to the minimum
description length (MDL) principle \cite{Wax85}. Then, the ESPRIT
(Estimation of Signal Parameters by Rotational Invariance Techniques) \cite%
{Roy1989} algorithm is employed in the second and third steps to detect
which codes are actually active and determine their corresponding timing
errors and CFOs.


\section{System description and signal model}

\subsection{System description}

We consider an OFDMA\ system employing $N$ subcarriers with index set $%
\{0,1,\ldots ,N-1\}$. As in \cite{Minn07}, we assume that a ranging
time-slot is composed by $M$ consecutive OFDMA\ blocks where the $N$
available subcarriers are grouped into ranging subchannels and data
subchannels. The former are used by the active RSSs to complete their
ranging processes, while the latter are assigned to data subscriber stations
(DSSs) for data transmission. We denote by $R$ the number of ranging
subchannels and assume that each of them is divided into $Q$ subbands. A
given subband is composed of a set of $V$ adjacent subcarriers which is
called a \textit{tile}. The subcarrier indices of the $q$th tile $%
(q=0,1,\ldots ,Q-1)$ in the $r$th subchannel $(r=0,1,\ldots ,R-1)$ are
collected into a set $\mathcal{J}_{q}^{(r)}=\{i_{q}^{(r)}+v\}_{v=0}^{V-1}$,
where the tile index $i_{q}^{(r)}$ can be chosen adaptively according to the
actual channel conditions. The only constraint in the selection of $%
i_{q}^{(r)}$ is that different tiles must be disjoint, i.e., $\mathcal{J}%
_{q_{1}}^{(r_{1})}\cap \mathcal{J}_{q_{2}}^{(r_{2})}=\varnothing $ for $%
q_{1}\neq q_{2}$ or $r_{1}\neq r_{2}$. The $r$th ranging subchannel is thus
composed of $QV$ subcarriers with indices in the set $\mathcal{J}^{(r)}=\cup
_{q=0}^{Q-1}{\mathcal{J}_{q}^{(r)}}$, while a total of $N_{R}=QVR$ ranging
subcarriers is available in each OFDMA block.

We assume that each subchannel can be accessed by a maximum number of $K_{\max }=\min \{V,M\}-1$
RSSs, which are separated by means of orthogonal codes in both the time and
frequency domains. The codes are selected in a pseudo-random fashion from a
predefined set $\{\mathbf{C}_{0},\mathbf{C}_{1},\ldots ,\mathbf{C}_{K_{\max
}-1}\}$ with
\begin{equation}
\mathbf{[C}_{k}\mathbf{]}_{v,m}=e^{j2\pi k(\frac{v}{V-1}+\frac{m}{M-1})}
\label{eq1}
\end{equation}%
where $v=0,1,\ldots ,V-1$ counts the subcarriers within a tile and is used
to perform spreading in the frequency domain, while $m=0,1,\ldots ,M-1$
is the block index by which spreading is done in the time domain across
the ranging time-slot. As in \cite{Minn07}, we assume that different RSSs
select different codes. Also, we assume that a selected code is employed by
the corresponding RSS over all tiles in the considered subchannel. Without
loss of generality, we concentrate on the $r$th subchannel and denote by $%
K\leq K_{\max }$ the number of simultaneously active RSSs. To simplify the
notation, the subchannel index $^{(r)}$ is dropped henceforth.

The signal transmitted by the $k$th RSS propagates through a multipath
channel characterized by a channel impulse response (CIR) $\mathbf{h}'%
_{k}=[h'_{k}(0),h'_{k}(1),\ldots ,h'_{k}(L-1)]^{T}$ of length $L$ (in sampling
periods). We denote by $\theta _{k}$ the timing error of the \textit{k}th
RSS expressed in sampling intervals $T_{s}$, while $\varepsilon _{k}$ is the
frequency offset normalized to the subcarrier spacing. As discussed in \cite%
{Morelli2004}, during IR the CFOs are only due to Doppler shifts and/or
to estimation errors and, in consequence, they are assumed to lie \textit{%
within a small fraction} of the subcarrier spacing. Timing offsets depend on
the distance of the RSSs from the BS and their maximum value is thus limited
to the round trip delay from the cell boundary. In order to eliminate
interblock interference (IBI), we assume that during the ranging process the
CP length comprises $N_{G}\geq \theta _{\max }+L$ sampling periods, where $%
\theta _{\max }$ is the maximum expected timing error. This assumption is
not restrictive as initialization blocks are usually preceded by long CPs in
many standardized OFDMA systems.

\subsection{System model}

We denote by $\mathbf{X}_{m}(q)=[X_{m}(i_{q}),X_{m}(i_{q}+1),\ldots
,X_{m}(i_{q}+V-1)]^{T}$ the discrete Fourier transform (DFT) outputs
corresponding to the $q$th tile in the $m$th OFDMA block. Since DSSs have
successfully completed their IR processes, they are perfectly aligned to the
BS references and their signals do not contribute to $\mathbf{X}_{m}(q)$. In
contrast, the presence of uncompensated CFOs destroys orthogonality among
ranging signals and gives rise to interchannel interference (ICI). The
latter results in a disturbance term plus an attenuation of the useful
signal component. To simplify the analysis, in the ensuing discussion the
disturbance term is treated as a zero-mean Gaussian random variable while
the signal attenuation is considered as part of the channel impulse
response. Under the above assumptions, we may write
\begin{equation}
X_{m}(i_{q}+v)=\sum\limits_{k=1}^{K}\mathbf{[C}_{\ell _{k}}\mathbf{]}_{v,m}{%
e^{j{m\omega _{k}N_{T}}}}H_{k}(\theta _{k},{\varepsilon _{k},}%
i_{q}+v)+w_{m}(i_{q}+v)  \label{eq2}
\end{equation}%
where $\omega _{k}=2\pi \varepsilon _{k}/N$, $N_{T}=N+N_{G}$ denotes the
duration of the cyclically extended block and $\mathbf{C}_{\ell _{k}}$ is
the code matrix selected by the \textit{k}th RSS. The quantity $H_{k}(\theta
_{k},{\varepsilon _{k},}n)$ is the $k$th \textit{equivalent} channel
frequency response over the $n$th subcarrier and is given by
\begin{equation}
H_{k}(\theta _{k},{\varepsilon _{k},}n)=\gamma _{N}({\varepsilon _{k})}%
H_{k}^{\prime }(n)e^{-j{2\pi n\theta _{k}}/N}  \label{eq3}
\end{equation}%
where $H_{k}^{\prime }(n)=\sum_{\ell =0}^{L-1}h_{k}^{\prime }(\ell )e^{-j{%
2\pi n\ell }/N}$ is the true channel frequency response, while
\begin{equation}
\gamma _{N}({\varepsilon })=\frac{\sin (\pi {\varepsilon })}{N\sin (\pi {%
\varepsilon }/N)}e^{j\pi {\varepsilon }(N-1)/N}  \label{eq5}
\end{equation}%
is the attenuation factor induced by the CFO. The last term in (\ref{eq2})
accounts for background noise plus interference and is modeled as a circularly symmetric
complex Gaussian random variable with zero-mean and variance $\sigma
_{w}^{2}=\sigma _{n}^{2}+\sigma _{ICI}^{2}$, where $\sigma _{n}^{2}$ and $%
\sigma _{ICI}^{2}$ are the average noise and ICI powers, respectively. From (%
\ref{eq3}) we see that $\theta _{k}$ appears only as a phase shift across
the DFT outputs. The reason is that the CP duration is longer than the
maximum expected propagation delay.

To proceed further, we assume that the tile width is much smaller than the
channel coherence bandwidth. In this case, the channel response is nearly
flat over each tile and we may reasonably replace the quantities $%
\{H_{k}^{\prime }(i_{q}+v)\}_{v=0}^{V-1}$ with an average frequency response
\begin{equation}
\overline{H}_{k}^{\prime }(q)=\frac{1}{V}\sum_{v=0}^{V-1}H_{k}^{\prime
}(i_{q}+v).  \label{eq6}
\end{equation}%
Substituting (\ref{eq1}) and (\ref{eq3}) into (\ref{eq2}) and bearing in
mind (\ref{eq6}), yields
\begin{equation}
X_{m}(i_{q}+v)=\sum\limits_{k=1}^{K}{e^{j2\pi ({m\xi _{k}+v\eta }_{k}{)}}}%
S_{k}(q)+w_{m}(i_{q}+v)  \label{eq7}
\end{equation}%
where $S_{k}(q)=\gamma _{N}({\varepsilon _{k})}\overline{H}_{k}^{\prime
}(q)e^{-j{2\pi i}_{q}{\theta _{k}}/N}$ and we have defined the quantities
\begin{equation}
\xi _{k}=\frac{\ell _{k}}{M-1}+\frac{\varepsilon _{k}N_{T}}{N}  \label{eq8}
\end{equation}
and
\begin{equation}
\eta _{k}=\frac{\ell _{k}}{V-1}-\frac{\theta _{k}}{N}  \label{eq9}
\end{equation}%
which are referred to as the \textit{effective }CFOs and timing errors,
respectively.

In the following sections we show how the DFT outputs $\{X_{m}(i_{q}+v)\}$
can be exploited to identify the active codes and to estimate the
corresponding timing errors and CFOs.

\section{ESPRIT-based estimation}

\subsection{Determination of the number of active codes}

The first problem to solve is to determine the number $K$ of active codes
over the considered ranging subchannel. For this purpose, we collect the $%
(i_{q}+v)$th DFT outputs across all ranging blocks into an \textit{M}%
-dimensional vector $\mathbf{Y}(i_{q,v})=[X_{0}(i_{q}+v),X_{1}(i_{q}+v),%
\ldots ,X_{M-1}(i_{q}+v)]^{T}$ given by
\begin{equation}
\mathbf{Y}(i_{q,v})=\sum_{k=1}^{K}e^{j2\pi v\eta _{k}}S_{k}(q)\mathbf{e}%
_{M}(\xi _{k})+\mathbf{w}(i_{q,v})  \label{eq30}
\end{equation}%
where $\mathbf{w}(i_{q,v})=[w_{0}(i_{q}+v),w_{1}(i_{q}+v),\ldots
,w_{M-1}(i_{q}+v)]^{T}$ is Gaussian distributed with zero mean and
covariance matrix $\sigma _{w}^{2}\mathbf{I}_{M}$ while $\mathbf{e}_{M}(\xi
)=[1,e^{j2\pi \xi },e^{j4\pi \xi },\ldots ,e^{j2\pi (M-1)\xi }]^{T}$.

From the above equation, we observe that $\mathbf{Y}(i_{q,v})$
has the same structure as measurements of multiple uncorrelated sources from
an array of sensors. Hence, an estimate of $K$ can be obtained by performing
an eigendecomposition (EVD) of the correlation matrix $\mathbf{R}_{Y}=%
\mathrm{E}\{\mathbf{Y}(i_{q,v})\mathbf{Y}^{H}(i_{q,v})\}$. In practice,
however, $\mathbf{R}_{Y}$ is not available at the receiver and must be
replaced by some suitable estimate. One popular strategy to get an estimate
of $\mathbf{R}_{Y}$ is based on the forward-backward (FB) principle.
Following this approach, $\mathbf{R}_{Y}$ is replaced by
$
\hat{\mathbf{R}}_{Y}=\frac{1}{2}(\tilde{\mathbf{R}}_{Y}+\mathbf{J}\tilde{%
\mathbf{R}}_{Y}^{T}\mathbf{J})$, where $\tilde{\mathbf{R}}_{Y}$ is the sample correlation matrix
\begin{equation}
\tilde{\mathbf{R}}_{Y}=\frac{1}{QV}\sum_{v=0}^{V-1}\sum_{q=0}^{Q-1}\mathbf{Y}%
(i_{q,v})\mathbf{Y}^{H}(i_{q,v})  \label{eq31}
\end{equation}%
while $\mathbf{J}$ is the exchange matrix with 1's on the anti-diagonal and
0's elsewhere. Arranging the eigenvalues $\hat{\lambda}_{1}\geq \hat{\lambda}%
_{2}\geq \cdots \geq \hat{\lambda}_{M}$ of $\hat{\mathbf{R}}_{Y}$ in
non-increasing order, we can find an estimate $\hat{K}$ of the number of
active codes by applying the MDL information-theoretic criterion. This
amounts to looking for the minimum of the following objective function \cite%
{Wax85}
\begin{equation}
\mathcal{F(}\tilde{K})=\frac{1}{2}\tilde{K}(2V-\tilde{K})\ln (MQ)-MQ(V-%
\tilde{K})\ln \rho (\tilde{K})  \label{eq22}
\end{equation}%
where $\rho (\tilde{K})$ is the ratio between the geometric and arithmetic
means of $\{\hat{\lambda}_{\tilde{K}+1},\hat{\lambda}_{\tilde{K}+2},\ldots ,%
\hat{\lambda}_{M}\}$.

\subsection{Frequency estimation}

For simplicity, we assume that the number of active codes has been perfectly
estimated. An estimate of ${\boldsymbol{\xi }}{=[}\xi _{1},\xi
_{2},\ldots ,\xi _{K}\mathbf{]}^{T}$ can be found by applying the ESPRIT
algorithm to the model (\ref{eq30}). To elaborate on this, we arrange the
eigenvectors of $\hat{\mathbf{R}}_{Y}$ associated to the $K$ largest
eigenvalues $\hat{\lambda}_{1}\geq \hat{\lambda}_{2}\geq \cdots \geq \hat{%
\lambda}_{K}$ into an $M\times K$ matrix $\mathbf{Z}=[\mathbf{z}_{1}~\mathbf{%
z}_{2}~\cdots \mathbf{z}_{K}]$. Next, we consider the matrices $\mathbf{Z}%
_{1}$ and $\mathbf{Z}_{2}$ that are obtained by collecting the first $M-1$
rows and the last $M-1$ rows of $\mathbf{Z}$, respectively. The entries of ${%
\boldsymbol{\xi }}$ are finally estimated in a decoupled fashion as
\begin{equation}
\hat{\xi}_{k}=\frac{1}{2\pi }\arg \{\rho _{y}(k)\}\text{ \ \ \ \ }%
k=1,2,\ldots ,K  \label{eq25}
\end{equation}%
where $\{\rho _{y}(1),\rho _{y}(2),\ldots ,\rho _{y}(K)\}$ are the
eigenvalues of
\begin{equation}
\mathbf{Z}_{Y}=(\mathbf{Z}_{1}^{H}\mathbf{Z}_{1})^{-1}\mathbf{Z}_{1}^{H}%
\mathbf{Z}_{2}  \label{eq26}
\end{equation}%
and $\arg \{\rho _{y}(k)\}$ denotes the phase angle of $\rho _{y}(k)$ taking
values in the interval $[-\pi ,\pi )$.

After computing estimates of the effective CFOs through (\ref{eq25}), the
problem arises of matching each $\hat{\xi}_{k}$ to the corresponding code $%
\mathbf{C}_{\ell _{k}}$. This amounts to finding an estimate of $\ell _{k}$
starting from $\hat{\xi}_{k}$. For this purpose, we denote by $\left\vert
\varepsilon _{\max }\right\vert $ the magnitude of the maximum expected CFO
and observe from (\ref{eq8}) that $(M-1)\xi _{k}$ belongs to the interval $%
I_{\ell _{k}}=[\ell _{k}-\beta ;\ell _{k}+\beta ]$, with $\beta =\left\vert
\varepsilon _{\max }\right\vert N_{T}(M-1)/N$. It follows that the effective
CFOs can be univocally mapped to their corresponding codes as long as $\beta
<1/2$ since only in that case intervals $\{I_{\ell _{k}}\}_{k=1}^{K}$ are
disjoint. The acquisition range of the frequency estimator is thus limited
to $\left\vert \varepsilon _{\max }\right\vert <N/(2N_{T}(M-1))$ and an estimate
of the pair $(\ell _{k},\varepsilon _{k})$ is computed as
\begin{equation}
\hat{\ell}_{k}=\text{round}\left((M-1)\hat{\xi}_{k}\right)  \label{eq35}
\end{equation}%
and
\begin{equation}
\hat{\varepsilon}_{k}=\frac{N}{N_{T}}\left(\hat{\xi}_{k}-\frac{\hat{\ell}_{k}}{M-1}\right).
\label{eq36}
\end{equation}%
It is worth noting that the $\arg \{\cdot \}$ function in (\ref{eq25}) has
an inherent ambiguity of multiples of $2\pi $, which translates into a
corresponding ambiguity of the quantity $\hat{\ell}_{k}$ by multiples of $M-1$%
. Hence, recalling that $\ell _{k}\in \{0,1,\ldots ,K_{\max }-1\}$ with $%
K_{\max } < M$, a refined estimate of $\ell _{k}$ can be found as
\begin{equation}
\hat{\ell}_{k}^{(F)}=[\hat{\ell}_{k}]_{M-1}  \label{eq29}
\end{equation}%
where $[x]_{M-1}$ is the value of $x$ reduced to the interval $[0,M-2]$. In
the sequel, we refer to (\ref{eq36}) as the ESPRIT-based frequency estimator
(EFE).

\subsection{Timing estimation}

We call $\mathbf{X}_{m}(q)=[X_{m}(i_{q}),X_{m}(i_{q}+1),\ldots
,X_{m}(i_{q}+V-1)]^{T}$ the $V$-dimensional vector of the DFT outputs
corresponding to the \textit{q}th tile in the \textit{m}th OFDMA\ block.
Then, from (\ref{eq7}) we have
\begin{equation}
\mathbf{X}_{m}(q)=\sum\limits_{k=1}^{K}{e^{j2\pi {m\xi _{k}}}}S_{k}(q)%
\mathbf{e}_{V}(\eta _{k})+\mathbf{w}_{m}(q)  \label{eq18}
\end{equation}%
where $\mathbf{w}_{m}(q)=[w_{m}(i_{q}),w_{m}(i_{q}+1),\ldots
,w_{m}(i_{q}+V-1)]^{T}$ is Gaussian distributed with zero mean and
covariance matrix $\sigma _{w}^{2}\mathbf{I}_{V}$ while $\mathbf{e}_{V}(\eta
)=[1,e^{j2\pi \eta },e^{j4\pi \eta },\ldots ,e^{j2\pi (V-1)\eta }]^{T}$.
Since $\mathbf{X}_{m}(q)$ is a superposition of complex sinusoidal signals
with random amplitudes embedded in white Gaussian noise, an estimate of ${%
\boldsymbol{\eta }}{=[}\eta _{1},\eta _{2},\ldots ,\eta _{K}\mathbf{]}^{T}$
can still be obtained by resorting to the ESPRIT algorithm. Following the
previous steps, we first compute $\hat{\mathbf{R}}_{X}=\frac{1}{2}(\tilde{%
\mathbf{R}}_{X}+\mathbf{J}\tilde{\mathbf{R}}_{X}^{T}\mathbf{J})$ with
\begin{equation}
\tilde{\mathbf{R}}_{X}=\frac{1}{MQ}\sum_{m=0}^{M-1}\sum_{q=0}^{Q-1}\mathbf{X}%
_{m}(q)\mathbf{X}_{m}^{H}(q).  \label{eq32}
\end{equation}%
Then, we define a $V\times K$ matrix $\mathbf{U}=[\mathbf{u}_{1}~\mathbf{u}%
_{2}~\cdots \mathbf{u}_{K}]$ whose columns are the eigenvectors of $\hat{%
\mathbf{R}}_{X}$ associated to the $K$ largest eigenvalues. The effective
timing errors are eventually estimated as
\begin{equation}
\hat{\eta}_{k}=\frac{1}{2\pi }\arg \{\rho _{x}(k)\},\text{ \ \ \ \ }%
k=1,2,\ldots ,K  \label{eq33}
\end{equation}%
where $\{\rho _{x}(1),\rho _{x}(2),\ldots ,\rho _{x}(K)\}$ are the
eigenvalues of
\begin{equation}
\mathbf{U}_{X}=(\mathbf{U}_{1}^{H}\mathbf{U}_{1})^{-1}\mathbf{U}_{1}^{H}%
\mathbf{U}_{2}  \label{eq34}
\end{equation}%
while the matrices $\mathbf{U}_{1}$ and $\mathbf{U}_{2}$ are obtained by
collecting the first $V-1$ rows and the last $V-1$ rows of $\mathbf{U}$,
respectively.

The quantities $\{\hat{\eta}_{k}\}_{k=1}^{K}$ are eventually used to find
estimates $(\hat{\ell}_{k},\hat{\theta}_{k})$ of the associated ranging code
and timing error. To accomplish this task, we let $\alpha =\theta _{\max
}(V-1)/(2N)$. Then, recalling that $0\leq \theta _{k}\leq \theta _{\max }$, from
(\ref{eq9}) we see that $(V-1)\eta _{k}+\alpha $ falls into the range $I_{\ell
_{k}}=[\ell _{k}-\alpha ;\ell _{k}+\alpha ]$. If $\theta _{\max }<N/(V-1)$, the
quantity $\alpha $ is smaller than 1/2 and, in consequence, intervals $%
\{I_{\ell _{k}}\}_{k=1}^{K}$ are disjoint. In this case, there is only one
pair $(\ell _{k},\theta _{k})$ that results into a given $\eta _{k}$ and an
estimate of $(\ell _{k},\theta _{k})$ is found as
\begin{equation}
\hat{\ell}_{k}=\text{round}\left((V-1)\hat{\eta}_{k}+\alpha \right)  \label{eq27}
\end{equation}%
and
\begin{equation}
\hat{\theta}_{k}=N\left(\frac{\hat{\ell}_{k}}{V-1}-\hat{\eta}_{k}\right).  \label{eq28}
\end{equation}%
As done in Sect. 3.2, a refined estimate of $\ell _{k}$ is obtained in the
form

\begin{equation}
\hat{\ell}_{k}^{(f)}=[\hat{\ell}_{k}]_{V-1}.  \label{eq37}
\end{equation}%
In the sequel, we refer to (\ref{eq28}) as the ESPRIT-based timing estimator
(ETE).

\subsection{Code detection}

From (\ref{eq29}) and (\ref{eq37}) we see that two distinct estimates\ $\hat{%
\ell}_{k}^{(F)}$ and $\hat{\ell}_{k}^{(f)}$ are available at the receiver
for each code index $\ell _{k}$. These estimates are now used to decide
which codes are actually active in the considered ranging subchannel. For
this purpose, we define two sets $I^{(f)}=\{\hat{\ell}_{k}^{(f)}\}_{k=1}^{K}$
and $I^{(F)}=\{\hat{\ell}_{k}^{(F)}\}_{k=1}^{K}$ and observe that, in the
absence of any detection error, it should be $I^{(f)}=I^{(F)}=\{\ell
_{k}\}_{k=1}^{K}$. Hence, for any code matrix $\mathbf{C}_{m}$
we suggest the following detection strategy
\begin{equation}
\begin{array}{c}
\text{if }m\in I^{(f)}\cap I^{(F)}\Longrightarrow \mathbf{C}_{m}\text{ is
declared \textit{detected }} \\
\text{if }m\notin I^{(f)}\cap I^{(F)}\Longrightarrow \mathbf{C}_{m}\text{ is
declared \textit{undetected}}.%
\end{array}
\label{eq37bis}
\end{equation}%
In the downlink response message, the BS will indicate only
the detected codes while undetected RSSs must restart their ranging process.
In the sequel, we refer to (\ref{eq37bis}) as the ESPRIT-based code detector
(ECD).

\section{Numerical results}

The investigated system has a total of $N=1024$ subcarriers over an uplink
bandwidth of 3 MHz. The sampling period is $T_{s}=0.33$ $\mu s$,
corresponding to a subcarrier distance of $1/(NT_{s})=2960$ Hz. We assume
that $R=4$ subchannels are available for IR. Each subchannel is divided into
$Q=16$ tiles uniformly spaced over the signal spectrum at a distance of $%
N/Q=64$ subcarriers. The number of subcarriers in any tile is $V=4$ while $%
M=4$. The discrete-time CIRs have $L=12$ taps which are modeled as
independent and circularly symmetric Gaussian random variables with zero
means and an exponential power delay profile, i.e., $E\{\left\vert
h_{k}(\ell )\right\vert ^{2}\}=\sigma _{h}^{2}\cdot \exp (-\ell /12)$ for $%
\ell =0,1,\ldots ,11$, where $\sigma _{h}^{2}$ is chosen such that $%
E\{\left\Vert \mathbf{h}_{k}\right\Vert ^{2}\}=1$. Channels of different
users are statistically independent of each other and are kept fixed over an
entire time-slot. We consider a maximum propagation delay of $\theta _{\max
}=204$ sampling periods. Ranging blocks are preceded by a CP of length $%
N_{G}=256$. The normalized CFOs are uniformly distributed over the interval $%
[-\Omega ,\Omega ]$ and vary at each run. Recalling that the estimation
range of EFE is $\left\vert \varepsilon _{k}\right\vert <N/(2N_{T}(M-1))$, we
set $\Omega \leq 0.1$.

\begin{figure}[htb]
\centering
\centerline{\epsfig{figure=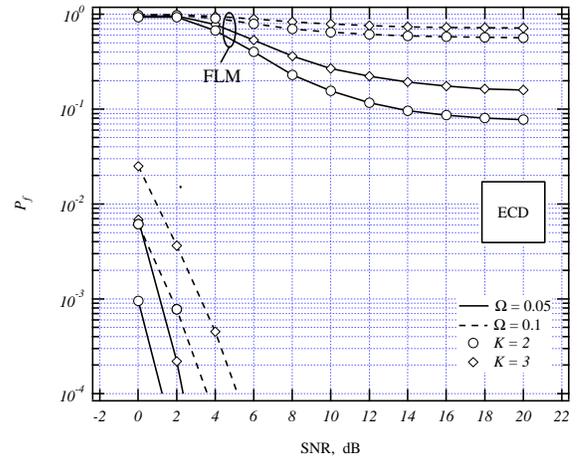,width=7.4cm}}
\caption{$P_{f}$ vs. SNR for $K$ = 3 when $\Omega$ is 0.05 or 0.1.}
\label{fig:res}
\end{figure}


\begin{figure}[htb]
\centering
\centerline{\epsfig{figure=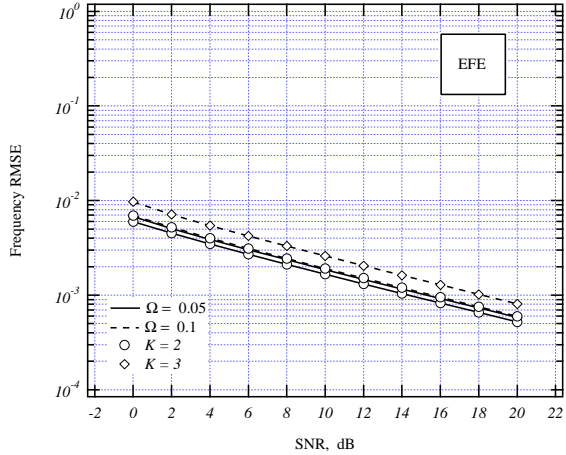,width=7.4cm}}
\caption{RMSE vs. SNR for $K$ = 2 or 3 when $\Omega$ is 0.05 or 0.1.}
\label{fig:res}
\end{figure}

\begin{figure}[htb]
\centering
\centerline{\epsfig{figure=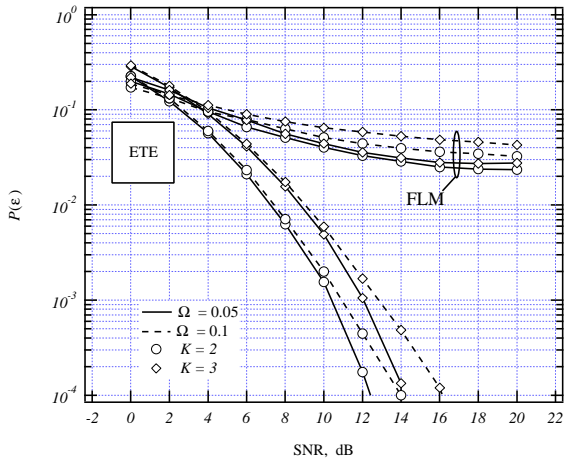,width=7.4cm}}
\caption{$P(\protect\epsilon)$ vs. SNR for $K$ = 2 or 3 when $\Omega$ is
0.05 or 0.1.}
\label{fig:res}
\end{figure}
We begin by investigating the performance of ECD in terms of probability of making an incorrect detection, say
$P_{f}$. Fig. 1 illustrates $P_{f}$ as a function of $%
\mathrm{SNR}=1/\sigma _{n}^{2}$. The number of active RSSs is 3 while
the maximum CFO is either $\Omega =0.05$ or 0.1. Comparisons are made with
the ranging scheme proposed by Fu, Li and Minn (FLM) in \cite{Minn07}. The
results of Fig. 1 indicate that ECD performs remarkably better than FLM.

Fig. 2 illustrates the root mean-square error (RMSE) of the frequency
estimates obtained with EFE vs. SNR for $K=2$ or 3 and $\Omega =0.05$ or
0.1. We see that the accuracy of EFE is satisfactory at SNR values
of practical interest. Moreover, EFE has virtually the same performance
as $K$ or $\Omega$ increase.

The performance of the timing estimators is measured in terms of probability
of making a timing error, say $P(\epsilon )$, as defined in \cite%
{Morelli2004}. An error event is declared to occur whenever the estimate $%
\hat{\theta}_{k}$ gives rise to IBI during the data section of the frame. This
is tantamount to saying that the timing
error $\hat{\theta}_{k}-\theta _{k}+(-N_{G,D}+L)/2$ is larger than zero or
smaller than $-N_{G,D}+L-1$, where $N_{G,D}$ is the CP length during the data transmission phase. In the sequel, we set $N_{G,D}=32$.
Fig. 3 illustrates $P(\epsilon )$ vs. SNR as
obtained with ETE and FLM. The number of active codes is $K=2$ or 3 while $%
\Omega =0.05$ or 0.1. We see that ETE provides much better results than FLM.

\section{Conclusions}

We have presented a new ranging method for OFDMA systems in which uplink
signals arriving at the BS are impaired by frequency errors in addition to
timing misalignments. The synchronization parameters of all ranging users
are estimated with affordable complexity through an ESPRIT-based approach.
Compared to previous works, the proposed scheme exhibits increased
robustness against residual frequency errors and can cope with situations
where the CFOs are as large as 10\% of the subcarrier spacing.

\bibliographystyle{IEEEbib}
\bibliography{IEEEabrv,../../../../../Bibliography/bibnew}


\end{document}